\documentclass[sigconf]{acmart}
\AtBeginDocument{%
  \providecommand\BibTeX{{%
    \normalfont B\kern-0.5em{\scshape i\kern-0.25em b}\kern-0.8em\TeX}}}

\copyrightyear{2023} 
\acmYear{2023} 
\setcopyright{acmlicensed}\acmConference[WWW '23]{Proceedings of the ACM Web Conference 2023}{April 30-May 4, 2023}{Austin, TX, USA}
\acmBooktitle{Proceedings of the ACM Web Conference 2023 (WWW '23), April 30-May 4, 2023, Austin, TX, USA}
\acmPrice{15.00}
\acmDOI{10.1145/3543507.3583339}
\acmISBN{978-1-4503-9416-1/23/04}

\usepackage{algorithm}
\usepackage{algorithmic}
\usepackage{makecell,booktabs}
\usepackage{multirow}
\usepackage{multicol}  
\usepackage{color}
\usepackage{hyperref}
\usepackage{enumitem}
\usepackage{subfigure}
\usepackage{xspace}

\newcommand{\name}{AutoDenoise\xspace}
\begin{document}

\title{\name: Automatic Data Instance Denoising for Recommendations}

\makeatletter
\def\@ACM@checkaffil{
    \if@ACM@instpresent\else
    \ClassWarningNoLine{\@classname}{No institution present for an affiliation}%
    \fi
    \if@ACM@citypresent\else
    \ClassWarningNoLine{\@classname}{No city present for an affiliation}%
    \fi
    \if@ACM@countrypresent\else
        \ClassWarningNoLine{\@classname}{No country present for an affiliation}%
    \fi
}
\makeatother
\author{Weilin Lin}
\affiliation{%
  \institution{City University of Hong Kong}
}
\email{mlmr.lin@gmail.com}

\author{Xiangyu Zhao}
\authornote{Xiangyu Zhao is the corresponding author.}
\affiliation{%
  \institution{City University of Hong Kong}
 }
 \email{xianzhao@cityu.edu.hk}
 
\author{Yejing Wang}
\affiliation{%
  \institution{City University of Hong Kong}
}
\email{yejing.wang@my.cityu.edu.hk}

\author{Yuanshao Zhu}
\affiliation{%
  \institution{City University of Hong Kong}
  \institution{Southern University of Science and Technology}
}
\email{zhuys2019@mail.sustech.edu.cn}
\author{Wanyu Wang}
\affiliation{%
  \institution{City University of Hong Kong}
  \country{}
 }
 \email{wanyuwang4-c@my.cityu.edu.hk}
\renewcommand{\shortauthors}{Weilin Lin, et al.}
\newcommand{\zxy}[1]{{\color{blue} [xiangyu: #1]}}
\newcommand{\w}[1]{{\color{red}[wyj:#1]}}
\newcommand{\lin}[1]{{\color{brown}[lwl:#1]}}
\newcommand{\zys}[1]{{\color{green}[ zys:#1]}}
\begin{abstract}
Historical user-item interaction datasets are essential in training modern recommender systems for predicting user preferences. 
However, the arbitrary user behaviors in most recommendation scenarios lead to a large volume of noisy data instances being recorded, which cannot fully represent their true interests.
While a large number of denoising studies are emerging in the recommender system community, all of them suffer from highly dynamic data distributions.  
In this paper, we propose a Deep Reinforcement Learning (DRL) based framework, \name, with an Instance Denoising Policy Network, for denoising data instances with an instance selection manner in deep recommender systems. 
To be specific, \name serves as an agent in DRL to adaptively select noise-free and predictive data instances, which can then be utilized directly in training representative recommendation models. 
In addition, we design an alternate two-phase optimization strategy to train and validate the \name properly.  
In the searching phase, we aim to train the policy network with the capacity of instance denoising; 
in the validation phase, we find out and evaluate the denoised subset of data instances selected by the trained policy network, so as to validate its denoising ability.
We conduct extensive experiments to validate the effectiveness of \name combined with multiple representative recommender system models.
\end{abstract}


\maketitle
\section{Introduction}
\label{sec:introduction}

Recommender System (RS) is an essential technology in the era of information explosion, which can deliver a favorable experience for users and considerable economic benefits for companies~\cite{kumar2014survey, gharibshah2020deep, lin2019cross}. 
As a data-driven technique, RS models user preference based on previous data instances (i.e., interaction logs)~\cite{ricci2011introduction}, thereby substantially improving the quality of online service platforms (e.g., e-commerce sites and streaming video applications)~\cite{wang2013opportunity, davidson2010youtube}.
In General, RS is modeled based on recorded data instances from historical user-item interactions. 
However, recording user behavior with the above approach inevitably raises challenges for modeling RS.
First, the existing recommendation dataset collects all historical data instances indiscriminately, while some of them may be noisy instances that do not reflect the real intention of users~\cite{hu2008collaborative}. 
For example, in \textit{click-through rate} (CTR) prediction~\cite{guo2017deepfm,wang2017deep,zhou2018deep,cheng2016wide, lin2022adafs, wang2022autofield, zhao2022adaptive, zhao2021autoloss}, some click actions may come from user curiosity or mistake.
Furthermore, the noisy instances have almost the same traits as noise-free instances~\cite{lee2021bootstrapping}, which renders a non-trivial task for human experts to separate them in the recording.

\begin{figure*}[t]
	\centering
	\Description{Show the whole framework with two proposed phases.}
	\includegraphics[width=0.66\linewidth]{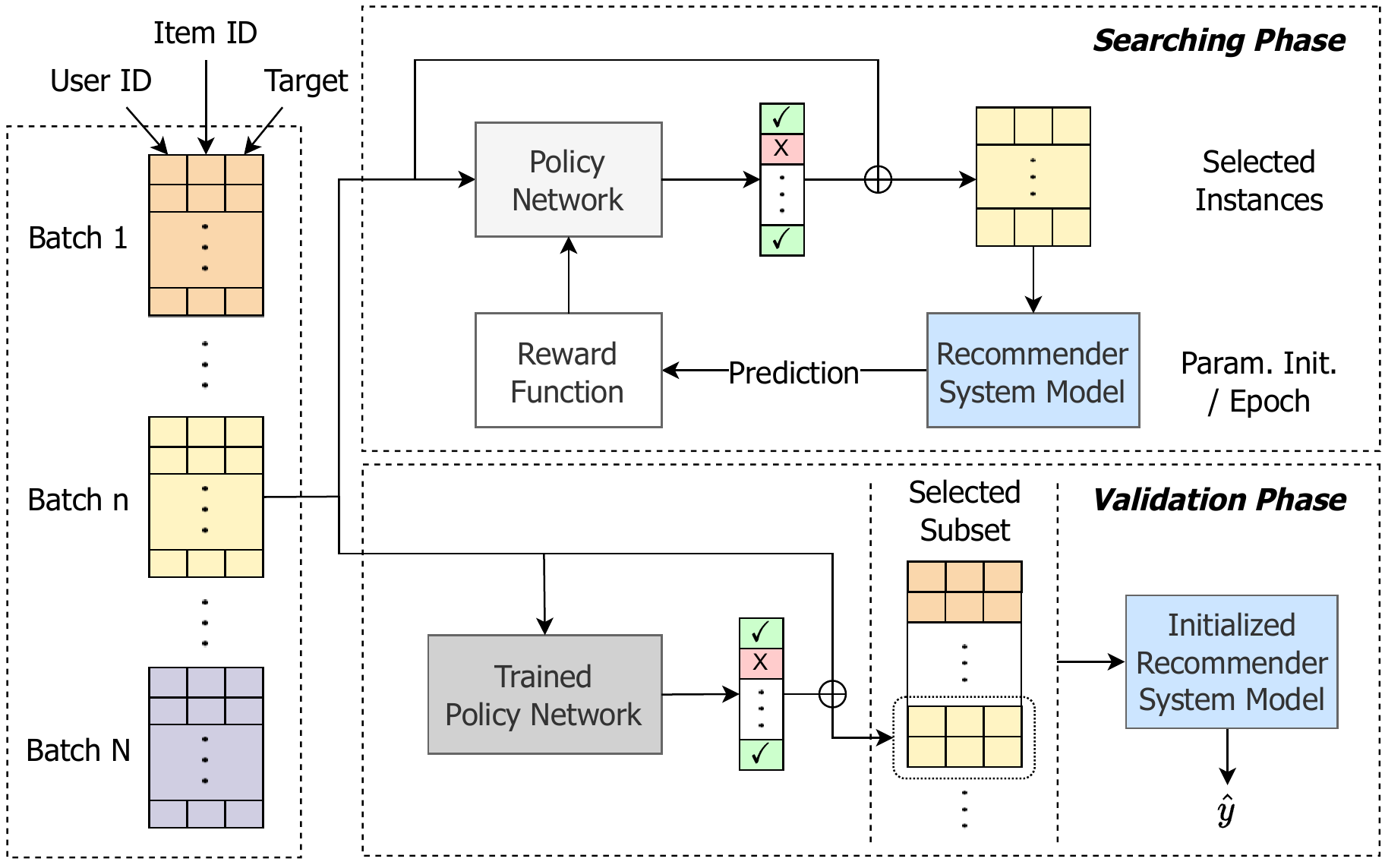}
	\vspace{-3mm}
	\caption{Overview of the proposed framework.}
	\label{fig:Fig2_overview}
\vspace{-3mm}
\end{figure*}

To address the above challenges, denoising training data is attracting growing attention from researchers.
Recent studies have shown that by using denoising techniques in recommender systems, models can be trained in a more efficient manner with better performance within comparable computational cost~\cite{gantner2012personalized, wang2021implicit,hu2021next, zhang2022hierarchical}.
Typically, denoising methods involve a ``searching'' behavior to figure out noise and a ``deciding'' behavior to execute denoising actions.
According to the methods of denoising, existing efforts can be roughly categorized into two groups: selection-based and reweighting-based methods.
For selection-based methods~\cite{ding2019sampler, yu2020sampler}, they aim to train a selection network to discard noisy instances, thereby feeding the model with more representative data.
For reweighting-based methods~\cite{ wang2021denoising, wang2022learning}, they tend to lower the contributions of the noisy instances by assigning lower weights throughout the model training. 
In practice, they consider the loss values as the indicator to discriminate noise from noise-free instances, i.e., high values imply noisy data instances during training.

However, the above methods suffer from the following limitations.
First, selection-based methods rely heavily on sampling distributions for decision-making~\cite{yuan2018fbgd}, while the complex and dynamic data distribution in the contemporary online environment tends to limit their performance.
In other words, the more dynamic environment is, the more biased selection may occur. 
Second, the reweighting-based methods influencing the model training processes lack transferability. 
Existing reweighting-based methods require specific configurations in the case of a given model or recommendation task, which is time-consuming and challenging to transfer to other settings.
Therefore, a promising instance denoising scheme should be robust and easily-transfer.

To bridge this research gap, we propose a Deep Reinforcement Learning (DRL) based instance denoising framework, which can model various complex data distributions of real-world datasets and filter a noise-free subset without noise.
We propose this framework based on three main motivations:
(i) The reinforcement learning methods are empirically proven to be effective in the optimum-searching problems~\cite{liu2019automating,liu2020automated, zhao2018deep, zhao2018recommendations, zhao2017deep, zhao2019deep, zhao2021dear}, which can also have the potential to distinguish the noisy data instances effectively. 
(ii) Policy network in DRL is used to select instances from a collection of mini-batch, allowing it to capture fine-grained patterns from various data distributions. 
Such design exactly facilitates mitigating the biased selection problem of selection-based methods.
(iii) Separately training the prediction model and the policy network for the denoising  and the prediction processes, owns natural transferability with the noise-free data subset from the denoising policy network.

Nevertheless, applying the DRL-based approach to instance denoising will encounter the following challenges.
On the one hand, the data distribution varies dramatically among mini-batches, rendering it challenging to design an appropriate reward function to optimize the policy network.
What's more, since the policy network and RS model need to be optimized simultaneously in this setting, it is challenging to train them properly with the same data batch.
To address the above challenges, we propose a DRL-based framework, \name, with an instance denoising policy network. 
\name scores each instance of a mini-batch by two probabilities for ``select/deselect'' actions with the policy network, and evaluates the performance of the sampled instances as ``select'' with the RS model. 
Meanwhile, we design an instance-level reward function based on the results of the RS model.
Specifically, this reward function compares a data instance's current loss with its losses in previous searching epochs, for learning to distinguish the noisy data instances in different distributions.
Moreover, we propose an alternate two-phase optimization strategy, i.e., the searching phase and the validation phase, to properly train and validate the policy network.
After that, we train a randomly initialized RS model on the selected data subset from scratch to evaluate the effectiveness of the new data subset. 
The main contributions of this work can be summarized as follows:
(i) We propose a DRL-based instance denoising framework, which adaptively filters out noises in each input mini-batch of data instances. To the best of our knowledge, this is a pioneering effort in instance denoising for CTR prediction;
(ii) We design a novel two-phase training process to effectively train and validate the policy network, as well as generate a transferable noise-free data subset;
(iii) We validate the effectiveness of \name on three public benchmark datasets and prove the transferability of the denoised datasets.

\begin{figure*}[t]
	\centering
	\Description{Show the whole process of selection}
	\includegraphics[width=0.9\linewidth]{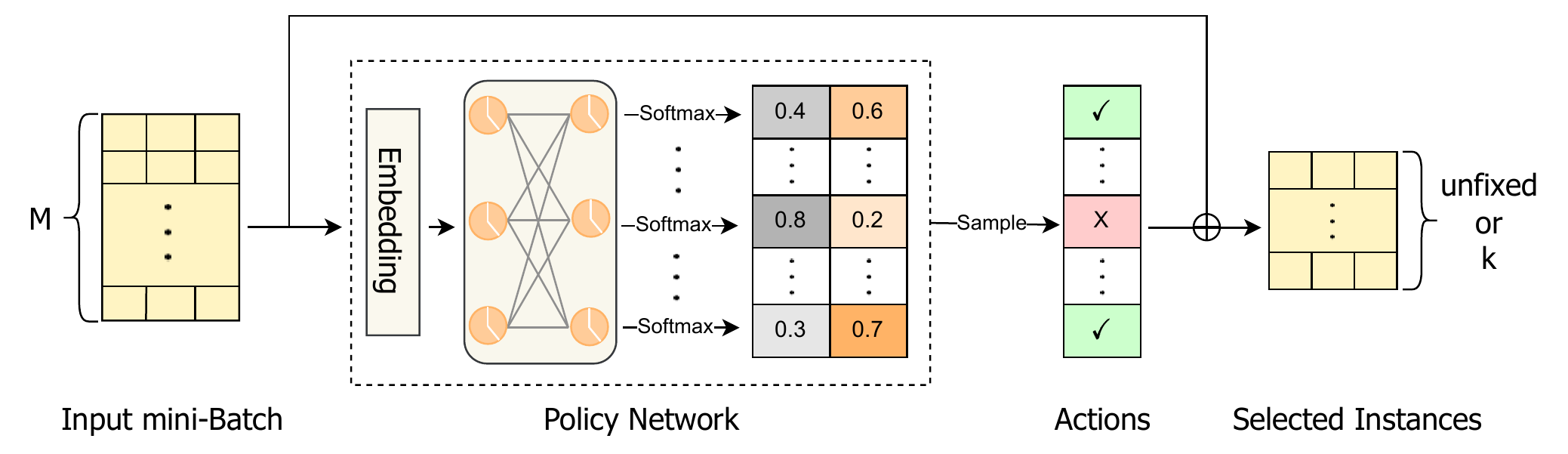}
	\vspace{-2.1mm} 
	\caption{The selection procedure of the policy network.}
	\label{fig:Fig3_policy}
\vspace{-3mm}
\end{figure*}

\section{Framework}
\label{sec:framework}

In this section, we will introduce the overview, the proposed two phases, and the optimization methods of \name framework.

\subsection{Overview}\label{subsec:overview}
As illustrated in Figure~\ref{fig:Fig2_overview}, the whole framework consists of two running phases, i.e., the searching phase and the validation phase.

\textbf{Searching Phase.} 
In this phase, DRL is adopted to automatically select instances with sequential optimization for the model and policy network.
Specifically, the policy network serves as a DRL agent, which performs ``select'' and ``deselect'' actions with corresponding probabilities, thereby removing noisy instances.
Then, the selected noise-free instances are fed into an RS model,
and the reward can be calculated by the previous and current losses of the RS model, which are used to optimize the policy network.

\textbf{Validation Phase.} Unlike the searching phase, this phase aims to evaluate the performance of the policy network. 
In detail, after finishing the searching phase of each epoch, i.e., one iteration on the entire training set, the process switches to the validation phase.
In the validation phase, the policy network predicts the entire training set in batches and selects the top-$k$ instances with the highest probability of ``select'' action, which can be considered as noise-free data.
Finally, an initialized RS model is trained with a noise-free subset until convergence, for which the final test performance is presented as the policy network denoising capability.

\subsection{Searching Phase}
\label{subsec:training} 
In this section, we focus on introducing core elements of policy networks and DRL construction (e.g., environment, state, rewards, etc.).
As an instance denoising task for applying DRL, we consider each mini-batch as the state of the input policy network and then execute the DRL actions (``select'' or ``deselect'' ) based on the outcome probabilities.
DRL interacts with the environment by feeding selected instances into the RS model and calculates the corresponding rewards, aiming to maximize the reward of the policy network.
Next, we will dive into the details of these components.

\textit{\textbf{Environment.}}
RS serves as the environment in DRL, which receives a mini-batch of selected instances and outputs the predictions. 
To fairly evaluate and optimize the policy network, we initialize its parameters at the beginning of each training epoch to make the output predictions comparable with previous ones.

\textit{\textbf{State.}}
The state is the input mini-batch, which consists of a bunch of data instances. Suppose that we have $M$ instances for a mini-batch and $N$ batches in total, the whole training set $\boldsymbol{X}=\left\{\boldsymbol{X}_1^n,...,\boldsymbol{X}_M^n\right\}_{n=1}^{N}$, 
where $\boldsymbol{X}_m^n$ represent the $m^{th}$ instance of the $n^{th}$ mini-batch in the training epoch. 
To compare the prediction results at the instance level for every training epoch, we fix their sequence order. 
In other words, all $\boldsymbol{X}_m^n$ have an identical position throughout the training process.
Since the state is the input mini-batch, the $n^{th}$ mini-batch can be expressed in the general form, i.e., $s_n=[ \boldsymbol{X}_1^n   \cdots   \boldsymbol{X}_m^n    \cdots    \boldsymbol{X}_M^n]$.
The function of the policy network is to score each instance in $s_n$ with probability and determine the ``select/deselect'' action.

\textit{\textbf{Policy and Action.}}
The policy and action can be considered as the ``searching'' and ``deciding'' behaviors in the instances of denoising, respectively.
Therefore, we propose a policy network to search for the noise-free instances based on the state $s_n$ and execute the action $a_m^n$ for each instance in the mini-batch $s_n$.
The whole selection process is illustrated in Figure~\ref{fig:Fig3_policy}. 
To be specific, given a batch of data $s_n = [\boldsymbol{X}_1^n \cdots \boldsymbol{X}_m^n \cdots \boldsymbol{X}_M^n]$, the policy network first transforms each instance $\boldsymbol{X}_m^n$ to a dense vector $\boldsymbol{E}_m^n$ through the embedding lookup operation, 
\begin{equation}
\begin{aligned}
&\boldsymbol{E}_m^n=\boldsymbol{\mathcal{A}}\boldsymbol{X}_m^n
\end{aligned}
\end{equation}
where $\boldsymbol{\mathcal{A}}$ consists of the learnable weight matrixes for all feature fields\footnote{A feature field is a group of feature values belonging to the same category, e.g., the feature field ``gender'' comprises two values, ``female'' and ``male''.}.
$\boldsymbol{E}_m^n$ denotes the concatenated feature embedding vector  corresponding to $\boldsymbol{X}_m^n$. 
Subsequently, $\boldsymbol{E}_m^n$ is fed to the Multilayer perceptron (MLP) with the nonlinear activation to obtain a fine-grained feature representation \cite{ramchoun2016multilayer}.
Suppose that the MLP has $\{L+1\}$ layers in total, the above operation of $l^{th}$ (${l} \in [1,{L}]$) layer can be formulated as:
\begin{equation}
\label{equ:mlp}
{\boldsymbol{h}_{l}=\sigma\left(\boldsymbol{W}_{l} \boldsymbol{h}_{l-1}+\boldsymbol{b}_{l}\right)}
\end{equation}
where $\boldsymbol{h}_{l}$ is the output of the ${l}^{th}$ layer.
$\boldsymbol{W}_{l}$ and $\boldsymbol{b}_{l}$ are the learnable weight matrix and bias vector for the corresponding layer, respectively, and $\sigma(\cdot)$ denotes the nonlinear activation function. 
In general, we use $ReLu$ as the activation function for hidden layers.
Then, we apply the $Softmax$ function to the output layer, so as to retrieve the probability of ``select'' and ``deselect'' actions for every instance.
In this setting, the action $a_m^n$ can be sampled from the output probabilities.
To reminder, $\boldsymbol{h}_{0} = \boldsymbol{E}_m^n$ represent the first fully-connected layer and the last layer $\boldsymbol{h}_{L}$ is the output layer.

\textit{\textbf{Reward.}}
To optimize the policy network in the desired way, i.e., learning to distinguish the noise instances, the reward design should be associated with the performance of the noise instance identification model. 
Following this idea, we define the reward according to the prediction error of the RS model, i.e., the value of loss $L$ comparing the prediction results and the corresponding ground-truth labels. 
In order to avoid optimization conflicts among data instances, we consider each instance separately in the training search phase. 
Specifically, we create a matrix 
$\boldsymbol{LMat} \in \mathbb{R}^{C \times MN}$ to store the loss values of previous and current searching epochs for each instance,
where $C$ and $MN$ are the numbers of stored epochs and the number of loss values. 
Given the loss $L_{mn}$ ($m^{th}$ instance of the $n^{th}$ batch) for the current epoch,
the reward $R_{mn}$ is defined as the difference between the corresponding instance's averaged loss values in the past $C$ epochs and the loss in the current epoch: 
\begin{equation}
\label{equ:reward}
R_{mn} = \frac{1}{C}\sum_{c=1}^C L_{mn}^c - L_{mn}
\end{equation}
where $R_{mn}>0$ means that the data instance's loss in the current epoch is smaller than its averaged loss values in the past $C$ epochs, indicating that the policy network conducts better action on the instance in the current epoch, and vice versa.
For fair comparison and improvement incentive, we initialize the parameters of the RS model at the beginning of each training epoch and overwrite the loss values of the earliest stored epoch with the latest one.
Under this setting, the reward can be viewed as the prediction improvement.
Since the parameters of the policy network are different, the reward can reflect the optimization results and be used to adjust them.
Considering the action-sampling behaviors of policy network would introduce randomness, which cannot represent its actual selection ability, we choose to use the average value of $L_{mn}^c$ in Equation~(\ref{equ:reward}) to make the current reward $R_{mn}$ more reliable.

\subsection{Validation Phase} 
\label{subsec:retrain}
As mentioned in the training phase, the RS model is initialized in each training epoch and usually not converged.
Therefore, we design a validation phase to fully evaluate the performance of the policy network and collect the best noise-free subset.
To implement this, we split the validation phase into noise-free subset selection and RS model training.
For subset selection, we iterate the entire training set and select the data instance with $k$ highest confidence in each mini-batch, i.e., the top-k selection strategy.
After obtaining the selected subset, we initialize the parameters of the RS model and train it until convergence, which is alignment with the typical recommendation system training setup.
Finally, with the same test and validation sets as in the searching phase, we can evaluate the quality of the selected subsets and the denoising ability of the trained policy network\footnote{In practice, we set a total of $50$ training epochs in the searching and validation phases to train and evaluate the policy network.}.
Only the policy network and noise-free subset with the best testing performance can be adopted.

{\it \textbf{Individual Selection \& Top-k Selection}}. 
Compared with the individual selection used in the searching phase, we adopt top-k selection in the validation phase, i.e., we select the $k$ prediction instances with the highest confidence by a trained policy network for each mini-batch.
The motivation for this scheme is individual selection provides a ``local'' view for policy network optimization, while the top-k can offer a ``global'' view enabling the selection of a more representative noise-free subset. 
From a policy training perspective, a ``local'' view at the instance level enables the policy network to learn the effectiveness of each instance and filter out noisy instances.
Nevertheless, the ``local'' view may not be the best choice for decision-making, because the contribution of weakly-selected instances may be minor compared to strongly-selected instances during model training, which may affect the model that uses the average of the loss values for optimization.
In this sense, weakly-selected instances can also be considered as a kind of noise, requiring a ``global'' constraint from the top-k selection strategy.

\subsection{Optimization Method}
\label{subsec:optimization}

With the working procedures mentioned above, we will further detail the optimization methods for updating both the RS model and the policy network in this section.

\subsubsection{Recommender System Model}
Since RS models in the searching phase and validation phase are identical, we introduce model parameters and the optimization method together. 
We denote the parameters of the RS model as $\boldsymbol{W}$. 
The CTR prediction could be regarded as a binary classification task under a supervised manner, where the input instances contain both features and the label. 
Thus, we apply \textit{binary-cross-entropy loss} function $\mathcal{L}_{model}$ to optimize $\boldsymbol{W}$:
\begin{equation}
\label{equ:loss_model}
\mathcal{L}_{model}(\boldsymbol{W})=-\frac{1}{M} \sum_{m=1}^{M} \left[ y_{m} \log \hat{y}_m + (1-y_{m}) \log (1-\hat{y}_m) \right]
\end{equation}
where $y_m$ and $\hat{y}_m$ are the ground truth label and the probability predicted by the RS model for the $m^{th}$ instance in a mini-batch, respectively. 
As we described in Section~\ref{subsec:training}, the loss value can be treated as the performance of the input selected instances in the searching phase. 
By minimizing the Equation~(\ref{equ:loss_model}), we obtain well-trained $\boldsymbol{W}$ as follows, where $\alpha$ is learning rate:
\begin{equation}
\label{equ:update_model}
\boldsymbol{W} \leftarrow \boldsymbol{W}-\alpha \nabla_{\boldsymbol{W}} \mathcal{L}_{model}(\boldsymbol{W})
\end{equation}

\begin{algorithm}[t]
	\caption{\label{alg:Opt_phaseI} $Opt\_I$: Optimization Algorithm for Searching Phase}
	\raggedright
	{\bf Input}: $N$ mini-Batches of data $\boldsymbol{X}=\left\{\boldsymbol{X}_1^n,...,\boldsymbol{X}_M^n\right\}_{n=1}^{N}$, policy network $\mathbf{\Theta}$, initial RS model $\boldsymbol{W}$, loss matrix $\boldsymbol{LMat}$\\
	{\bf Output}: trained policy network $\mathbf{\Theta}^*$ and full loss vector $\boldsymbol{LVec}^*$\\
	\begin{algorithmic} [1]
	\STATE Create empty $\boldsymbol{LVec}$
	\STATE Initialize $\boldsymbol{W}$
	\FORALL{$n \in [1,N]$} 
    \STATE $s_n \leftarrow [\boldsymbol{X}_1^n \cdots \boldsymbol{X}_m^n \cdots \boldsymbol{X}_M^n]$
    \STATE Sample $a_{1} \sim \mathbf{\Theta}(a \mid s_n)$
    \STATE Select instances $\boldsymbol{X}_{slct1}$ 
    \STATE Estimate $\mathcal{L}_{model}(\boldsymbol{W})$ by Equation~(\ref{equ:loss_model})
    \STATE Estimate $R_n$ by Equation~(\ref{equ:reward}) with $\boldsymbol{LMat}$
    \STATE Update $\mathbf{\Theta}$ by Equation~(\ref{equ:update_policy})
    \STATE Sample $a_{2} \sim \mathbf{\Theta}(a\mid s)$
    \STATE Select instances $\boldsymbol{X}_{slct2}$
    \STATE Estimate $\mathcal{L}_{model}(\boldsymbol{W})$ by Equation~(\ref{equ:loss_model})
    \STATE Append $\mathcal{L}_{model}(\boldsymbol{W})$ to $\boldsymbol{LVec}$
    \STATE Update $\boldsymbol{W}$ by Equation~(\ref{equ:update_model})
    \ENDFOR
    \STATE $\mathbf{\Theta}^* \leftarrow \mathbf{\Theta}$
    \end{algorithmic}
    \vspace{-0.8mm}
\end{algorithm}

\subsubsection{Policy Network}
As it is a reinforcement learning framework, the optimal policy network can be parameterized by $\mathbf{\Theta}$. 
Given the reward function, i.e., Equation~(\ref{equ:reward}), the objective function required optimization is formulated as:
\begin{equation}
\label{equ:objective_function}
\max _{\mathbf{\Theta}} J(\mathbf{\Theta})=\mathbb{E}_{a \sim \mathbf{\Theta}(a \mid s)} \left[R(a\mid s)\right]
\end{equation}
where $a \sim \mathbf{\Theta}(a \mid s)$ means an action $a$ is sampled via the policy $\mathbf{\Theta}(a \mid s)$ in state $s$.
To maximize the above function, we apply a policy gradient algorithm, named REINFORCE~\cite{williams1992simple}, and adopt Monte-Carlo sampling to simplify the estimation of the gradient $\nabla_{\mathbf{\Theta}} J(\mathbf{\Theta})$, which can be expressed as:
\begin{equation}
\begin{aligned}
\nabla_{\mathbf{\Theta}} J(\mathbf{\Theta}) &=\sum_{a} R(a \mid s) \nabla \mathbf{\Theta}(a \mid s) \\
&=\sum_{a} R(a \mid s) \mathbf{\Theta}(a \mid s) \nabla \log \mathbf{\Theta}(a \mid s) \\
&=\mathbb{E}_{a \sim \mathbf{\Theta}(a \mid s)}[R(a \mid s) \nabla \log \mathbf{\Theta}(a \mid s)] \\
& \approx \frac{1}{T} \sum_{t=1}^{T} R(a \mid s) \nabla \log \mathbf{\Theta}(a \mid s)
\end{aligned}
\end{equation}
where $T$ is the sample number, and we set $T=1$ here to improve the computational efficiency. 
The parameters $\mathbf{\Theta}$ can be updated as:
\begin{equation}
\label{equ:update_policy}
\mathbf{\Theta} \leftarrow \mathbf{\Theta}+\beta \nabla_{\mathbf{\Theta}} J(\mathbf{\Theta})
\end{equation}
where $\beta$ is the learning rate.

\begin{algorithm}[t]
	\caption{\label{alg:Opt_phaseII} $Opt\_II$: Optimization Algorithm for Validation Phase }
	\raggedright
	{\bf Input}: $N$ mini-Batches of data $\boldsymbol{X}=\left\{\boldsymbol{X}_1^n,...,\boldsymbol{X}_M^n\right\}_{n=1}^{N}$, trained policy network $\mathbf{\Theta}^*$, initial RS model $\boldsymbol{W}$ \\
	{\bf Output}: Selected data subset $\boldsymbol{X}_{sub}$, well-trained RS model $\boldsymbol{W}^*$\\
	\begin{algorithmic} [1]
	\STATE Create empty $\boldsymbol{X}_{sub}$
	\STATE Initialize $\boldsymbol{W}$
    \FORALL{$n \in [1,N]$} 
    \STATE $s_n \leftarrow [\boldsymbol{X}_1^n \cdots \boldsymbol{X}_m^n \cdots \boldsymbol{X}_M^n]$
    \STATE Sample $a_{3} \sim \mathbf{\Theta}^*(a \mid s_n)$
    \STATE Select instances $\boldsymbol{X}_{slct3}$
    \STATE Append $\boldsymbol{X}_{slct3}$ to $\boldsymbol{X}_{sub}$
    \ENDFOR
    \WHILE{RS model $\boldsymbol{W}$ not converged}
        \STATE Sample a batch of data from $\boldsymbol{X}_{sub}$
        \STATE Update $\boldsymbol{W}$ by Equation~(\ref{equ:update_model})
        \ENDWHILE
    \end{algorithmic}
    \vspace{-0.8mm}
\end{algorithm}

\subsubsection{Optimization Algorithm for Searching Phase}
Based on the optimization methods for the RS model and the policy network mentioned above, we can conduct the two-phase training to optimize and evaluate the proposed framework. 
The searching phase is illustrated in Algorithm~\ref{alg:Opt_phaseI}. 
In this phase, we iterate the whole training set and alternatively optimize $\mathbf{\Theta}$ and $\boldsymbol{W}$ to obtain a well-performed policy. 
For a fair comparison of the reward, we initialize the RS model $\boldsymbol{W}$ at the beginning of this phase (line 2). 
During the training epoch, we first sample a mini-batch of data instances as the state $s_n$ for the policy network (line 4). 
Next, we sample actions $a_1$ for every data instance in the batch (line 5) and collect them as a selective batch $\boldsymbol{X}_{slct1}$ (line 6), which will be fed into the RS model and gain the corresponding loss values (line 7). 
With this loss value and the saved ones from previous epochs\footnote{For the first $C$ epochs in searching phase, where previous $C$ losses are now available, we design a warm-up train method in Section \ref{sec:overallopt}.}, i.e., in $\boldsymbol{LMat}$, we can calculate the reward (line 8) and update our policy network (line 9). 
Similarly, the optimization of the RS model (line 10-14) follows a similar procedure as in line 5 to line 9. 
However, we sample new actions $a_2$ according to the updated $\mathbf{\Theta}$ (line 10), and update $\boldsymbol{W}$ (line 14) rather than $\mathbf{\Theta}$ as line 9. 
The updated policy network in the final mini-batch will be returned as a well-trained policy network (line 16). 
What's more, to record loss values for future calculation of reward baseline, we create an empty loss vector $\boldsymbol{LVec}$ (line 1) and record the test loss value (line 13) during the iteration. 

Specifically, the policy network $\mathbf{\Theta}$ from the previous optimization step cannot represent the optimized performance of the current batch of data.
We utilize every mini-batch of data twice in this phase to sequentially optimize $\mathbf{\Theta}$ and $\boldsymbol{W}$ (line 9 and 14), thus avoiding disturbing the returning loss and the optimizing environment.

\begin{algorithm}[t]
	\caption{\label{alg:Optimization} Overall Optimization Algorithm of \name}
	\raggedright
	{\bf Input}: $N$ mini-Batches of data $\boldsymbol{X}=\left\{\boldsymbol{X}_1^n,...,\boldsymbol{X}_M^n\right\}_{n=1}^{N}$, Warm-up epoch ${C}$, Training epoch $T$, empty loss matrix $\boldsymbol{LMat} = [\boldsymbol{LVec}_1,...,\boldsymbol{LVec}_C]$, initial RS model $\boldsymbol{W}$ and policy network $\mathbf{\Theta}$\\
	{\bf Output}: well-trained policy network parameters $\mathbf{\Theta}^*$ and the selected data subset $\boldsymbol{X}_{sub}$\\
	\begin{algorithmic} [1]
	    \FOR{$i = 1$; $i<={C}$; $i++$ } 
	    \STATE Initialize $\boldsymbol{W}$
	    \WHILE{RS model $\boldsymbol{W}$ not converged}
        \STATE Sample a batch of data $[\boldsymbol{X}_1^n \cdots \boldsymbol{X}_m^n \cdots \boldsymbol{X}_M^n]$
        \STATE Estimate $\mathcal{L}_{model}(\boldsymbol{W})$ by Equation~(\ref{equ:loss_model})
        \STATE Append $\mathcal{L}_{model}(\boldsymbol{W})$ to $\boldsymbol{LVec}_i$
        \STATE Update $\boldsymbol{W}$ by Equation~(\ref{equ:update_model})
        \ENDWHILE
	    \ENDFOR
	    
	    \FOR{$t = 1$; $t<=T$; $t++$} 
	   
	    \STATE $\mathbf{\Theta}^*, \boldsymbol{LVec}^* \leftarrow Opt\_I(\boldsymbol{X}, \mathbf{\Theta}, \boldsymbol{W}, \boldsymbol{LMat})$ 
	    \STATE $\boldsymbol{LVec}_{t\%C} \leftarrow \boldsymbol{LVec}^*$
	    
        \STATE $\boldsymbol{X}_{sub}, \boldsymbol{W}^* \leftarrow Opt\_II(\boldsymbol{X}, \mathbf{\Theta}^*, \boldsymbol{W})$
        \STATE Test $\boldsymbol{W}^*$
        \IF{$\boldsymbol{W}^*$ perform the best}
        \STATE Save $\boldsymbol{X}_{sub}$ and $\mathbf{\Theta}^*$
        \ENDIF
	    \ENDFOR
	\end{algorithmic}
 \vspace{-0.8mm}
\end{algorithm}
\subsubsection{Optimization Algorithm for Validation Phase}
In Algorithm~\ref{alg:Opt_phaseII}, we demonstrate the whole procedure of the validation phase, including the selection process (line 3-8) and the full-training process of the RS model (line 9-12). 
In this phase, we aim to evaluate the true performance of the trained policy network $\mathbf{\Theta}^*$ by training the RS model to converge on the selected subset. Specifically, we first create an empty set $\boldsymbol{X}_{sub}$ (line 1) and initialize $\boldsymbol{W}$ (line 2). 
Then, we iterate the whole training set for subset selection with $\mathbf{\Theta}^*$. 
Similar to the procedure in the searching phase, every mini-batch will follow the sampling (line 5) and selection (line 6) processes. 
Then the obtained noise-free data batch $\boldsymbol{X}_{slct3}$ is appended to $\boldsymbol{X}_{sub}$ (line 7).
After that, we will train the initialized $\boldsymbol{W}$ to converge with $\boldsymbol{X}_{sub}$ (line 9-12). 
And we consider the performance of the well-trained $\boldsymbol{W}^*$ as our validation result.

\begin{table*}[t]
\caption{Overall performance with different backbone recommendation models.}
\vspace{-3mm}
\label{tab:over1}
\begin{tabular}{@{}c|c|cc|cc|cc|cc|cc@{}}
\toprule
\multirow{2}{*}{\begin{tabular}[c]{@{}c@{}}Backbone \\ Model\end{tabular}} & \multirow{2}{*}{Metrics} & \multicolumn{2}{c|}{FM}   & \multicolumn{2}{c|}{Wide \& Deep} & \multicolumn{2}{c|}{DCN}  & \multicolumn{2}{c|}{IPNN} & \multicolumn{2}{c}{DeepFM}  \\ \cmidrule{3-12} 
                                                                           &                          & \textbf{w/o}    & \textbf{w}               & \textbf{w/o}        & \textbf{w}                   & \textbf{w/o}    & \textbf{w}               & \textbf{w/o}    & \textbf{w}               & \textbf{w/o}    & \textbf{w}               \\ 
\cmidrule{1-12}
\multirow{2}{*}{MovieLens1M}                                              
& AUC  $\uparrow$                    & 0.8080 & \textbf{0.8095*} & 0.7957     & \textbf{0.8022*}     & 0.8092 & \textbf{0.8115*} & 0.8023 & \textbf{0.8063*} & 0.8048 & \textbf{0.8097*} \\
& Logloss       $\downarrow$           & 0.5334 & \textbf{0.5292*} & 0.5400     & \textbf{0.5333*}     & 0.5235 & \textbf{0.5219*} & 0.5433 & \textbf{0.5363*} & 0.5306 & \textbf{0.5225*} \\ 
\cmidrule{1-12}
\multirow{2}{*}{KuaiRec-Small}                                             
& AUC    $\uparrow$                  & 0.8654 & \textbf{0.8657} & 0.8641     & \textbf{0.8642}     & 0.8641 & \textbf{0.8649} & 0.8630 & \textbf{0.8635} & 0.8653 & \textbf{0.8659} \\
& Logloss       $\downarrow$           & 0.4183 & \textbf{0.4178} & 0.4204     & \textbf{0.4201}     & 0.4202 & \textbf{0.4192*} & 0.4228 & \textbf{0.4217*} & 0.4184 & \textbf{0.4172*} \\ 
\cmidrule{1-12}
\multirow{2}{*}{Netflix}                                                  
& AUC    $\uparrow$                  & 0.6523 & \textbf{0.6541*} & 0.6575     & \textbf{0.6620*}     & 0.6711 & \textbf{0.6761*} & 0.6682 & \textbf{0.6734*} & 0.6604 & \textbf{0.6683*} \\
& Logloss      $\downarrow$            & 0.7835 & \textbf{0.7803*} & 0.9350     & \textbf{0.9115*}     & 0.7742 & \textbf{0.7434*} & 0.7760 & \textbf{0.7542*} & 0.8926 & \textbf{0.8800*} \\ 
\bottomrule
\end{tabular}
\\``\textbf{{\Large *}}'' indicates the statistically significant improvements (i.e., two-sided t-test with $p<0.05$) over the \textbf{w/o} version.
\vspace{-3mm}
\end{table*}
\begin{table*}[t]
\caption{Overall performance comparison with baselines. (Backbone model: DeepFM)}
\vspace{-3mm}
\label{tab:overall}
\begin{tabular}{@{}c|c|c|cccc|cc|c@{}}
\toprule
Dataset & Metrics & \textbf{w/o} & Drop3  & BDIS   & LSBo   & LSSm   & T-CE   & R-CE   & \name   \\ \midrule
\multirow{2}{*}{MovieLens1M}   & AUC $\uparrow$  & 0.8048 & 0.7701 & 0.6995 & 0.7659 & 0.7752 & 0.7814 & 0.7817 & \textbf{0.8097*} \\
        & Logloss $\downarrow$ & 0.5306 & 0.5776 & 0.7996 & 0.5851 & 0.5841 & 1.7886 & 1.8263  & \textbf{0.5225*} \\ \midrule
\multirow{2}{*}{KuaiRec-Small} & AUC $\uparrow$  & 0.8653 & 0.8485 & 0.8018 & 0.8439 & 0.8566 & 0.8140 & 0.8189 & \textbf{0.8659*} \\
        & Logloss $\downarrow$ & 0.4184 & 0.4472 & 0.6694 & 0.4501 & 0.4931 & 1.7509 & 1.8955  & \textbf{0.4172*} \\ \midrule
\multirow{2}{*}{Netflix}       & AUC $\uparrow$  & 0.6604 & 0.6444 & 0.5726 & 0.6355 & 0.6471 & 0.6785 & 0.6832  & 0.6683 \\
        & Logloss $\downarrow$ & 0.8926 & 1.3277 & 1.1192 & 1.4491 & 1.0330 & 2.5780 & 2.5801  & \textbf{0.8800*}\\ \bottomrule
\end{tabular}
\\``\textbf{{\Large *}}'' indicates the statistically significant improvements over the best baseline. $\uparrow$: higher is better; $\downarrow$: lower is better.
\vspace{-3mm}
\end{table*}

\subsubsection{Overall Optimization Algorithm}
\label{sec:overallopt}
With the optimization algorithms for the searching phase and validation phase, we will further detail the overall optimization algorithm in this subsection. The whole algorithm is depicted in Algorithm~\ref{alg:Optimization}.

{\it \textbf{Warm-up Train}}.
To build the loss matrix $\boldsymbol{LMat}$ for reward calculation, we conduct a Warm-up Train before the searching phase. 
Specifically, only the RS model will be trained under the same settings as in the searching phase, i.e., being initialized at the beginning of every epoch and using the same loss function, denoted as $\mathcal{L}_{model}$. 
After iterating $C$ epochs, we can obtain a $\{C \times MN\}$ matrix and start the searching phase with it.

The whole optimizing procedure consists of a warm-up train (line 1-9) and two training phases within the testing process (line 10-18). 
In the warm-up train, we first initialize the parameters of RS model $\boldsymbol{W}$ and train it to converge for $C$ epochs (line 1-9), where we sample a mini-batch (line 4) for model optimization (line 5) and save the corresponding loss values (line 6) for the reward calculation. 
After that, we follow Algorithm~\ref{alg:Opt_phaseI} (line 11) and Algorithm~\ref{alg:Opt_phaseII} (line 13) to train and evaluate the policy network.
With the $\boldsymbol{W}^*$ optimized by the selected subset, we can finally evaluate its performance (line 14) and save the optimal policy and the subset (line 15-17). 
The saved $\boldsymbol{X}_{sub}$ after the total $T$ epochs is considered to be the noise-free subset.
In the practical inference stage, we replace it with the original training set for model training.

\section{Experiment}\label{sec:experiment}
In this section, we conduct extensive experiments on three public datasets to investigate the effctiveness of \name.

\subsection{Experimental Settings}
\subsubsection{Datasets}
We evaluate the model performance of \name and baselines on three public datasets with different densities, \textbf{MovieLens1M\footnote{\url{https://grouplens.org/datasets/movielens/1m/}}} (4.47\%),
\textbf{KuaiRec-Small\footnote{\url{https://github.com/chongminggao/KuaiRec}}} (99.62\%), and \textbf{Netflix\footnote{\url{https://www.kaggle.com/datasets/netflix-inc/netflix-prize-data}}} (0.02\%).
For all datasets, we randomly select $80\%$ as the training set, $10\%$ as the validation set, and the remaining $10\%$ as a test set.

\subsubsection{Evaluation Metrics}
To fairly evaluate recommendation performance, we select the AUC (Area Under the ROC Curve) scores and Logloss (logarithm of the loss value) as metrics, which are widely used for click-through prediction tasks~\cite{guo2017deepfm,cheng2016wide}. 
In practice, higher AUC scores or lower logloss values at the 0.001-level indicate a significant improvement~\cite{cheng2016wide}.

\subsubsection{Implementation Details}
The implementation of \name is based on a PyTorch-based public library\footnote{https://github.com/rixwew/pytorch-fm}, which involves sixteen state-of-the-art RS models. We develop the policy network of \name as an individual class so that it can easily incorporate with different backbone RS models, like in Table~\ref{tab:overall}.
Specifically, we implement it as a two-layer MLP with additional embedding layer and output layer. Every fully-connected layer in the MLP is stacked with a linear layer, a batch normalization operation, a $ReLU$ activation function, and a dropout operation (\textit{rate = 0.2}). The output layer is implemented with a \textit{softmax} activation. For the hyper-parameter, we set both the embedding size in the embedding layer and the dimension size of the two fully-connected layers as $d=16$. The dimension of the output layer is set as 2, so as to generate two action probabilities for every instance.

\begin{table*}[ht]
\caption{Transferability test.}
\vspace{-3mm}
\label{tab:tran}
\begin{tabular}{c|c|cccccc}
\toprule
\multirow{2}{*}{\begin{tabular}[c]{@{}c@{}}Recommendation\\ Model\end{tabular}} &
\multirow{2}{*}{Methods} &  \multicolumn{2}{c}{MovieLens1M} &  
\multicolumn{2}{c}{KuaiRec-Small} &  \multicolumn{2}{c}{Netflix} \\\cline{3-8}
&        & AUC      $\uparrow$       & Logloss    $\downarrow$      & AUC     $\uparrow$         & Logloss    $\downarrow$      & AUC       $\uparrow$       & Logloss   $\downarrow$       \\\hline
\multirow{2}{*}{FM}           
& Normal & 0.8080          & 0.5334          & 0.8654          & 0.4183          & 0.6523          & 0.7835          \\
& \name   & \textbf{0.8082} & \textbf{0.5274*} & \textbf{0.8657} & \textbf{0.4179} & \textbf{0.6532*} & \textbf{0.7783*} \\\hline
\multirow{2}{*}{Wide \& Deep} 
& Normal & 0.7957          & 0.5400          & 0.8641          & 0.4204          & 0.6575          & 0.9350          \\
& \name   & \textbf{0.8001*} & \textbf{0.5363*} & \textbf{0.8645} & \textbf{0.4202} & \textbf{0.6587*} & \textbf{0.9212*} \\\hline
\multirow{2}{*}{DCN}          
& Normal & 0.8092          & 0.5235          & 0.8641          & 0.4202          & 0.6711          & 0.7742          \\
& \name   & \textbf{0.8111*} & \textbf{0.5230} & \textbf{0.8648} & \textbf{0.4193} & \textbf{0.6762*} & \textbf{0.7476*} \\\hline
\multirow{2}{*}{IPNN}         
& Normal & 0.8023          & 0.5433          & 0.8630          & 0.4228          & 0.6682          & 0.7760          \\
& \name   & \textbf{0.8029} & \textbf{0.5368*} & \textbf{0.8644*} & \textbf{0.4206*} & \textbf{0.6723*} & \textbf{0.7588*}
\\\bottomrule
\end{tabular}
\\``\textbf{{\Large *}}'' indicates the statistically significant improvements over the best baseline. $\uparrow$: higher is better; $\downarrow$: lower is better.
\vspace{-3mm}
\end{table*}

\subsection{Overall Performance} 
\label{sec:overall_performance}
This section evaluates the effectiveness of \name.
As in Table~\ref{tab:over1}, we compare the performances of recommendation models trained on a noise-free dataset denoised by \name (\textbf{w}), against models trained on the original dataset with noise (\textbf{w/o}), and then evaluate them on the identical test dataset.
We choose various advanced backbone recommendation models, including FM~\cite{rendle2010factorization}, Wide \& Deep~\cite{cheng2016wide}, DCN~\cite{wang2017deep}, and DeepFM~\cite{guo2017deepfm}. 
In addition, 
we also compare the denoising quality of \name with state-of-the-art instance selection methods (Drop3~\cite{wilson2000reduction}, BDIS~\cite{Chen_Cao_Xing_Liang_2022}, LSBo~\cite{leyva2015three}, LSSm~\cite{leyva2015three}) and instance denoising methods (T\_CE~\cite{wang2021denoising}, R\_CE~\cite{wang2021denoising})\footnote{These two baselines could be applied to general denoising tasks. We do not test the performance of other denoising models since the different task settings as discussed in Section 3.1.} in Table~\ref{tab:overall}.
We could conclude that: 
\begin{itemize}[leftmargin=*]
    \item In light of the result in Table~\ref{tab:over1}, integrating \name can boost the performance for all backbone models on all datasets. 
    Since KuaiRec-Small is far denser than MovieLens1M and Netflix (Data density: 99.62\% $\gg$ 4.47\% $>$ 0.02\%), the less remarkable improvement may come from the dominant contribution of clean instances, while noisy instances contribute less to user preference modeling, reasoning that the dense user-item interactions can accurately show the whole picture of a user. On the contrary, for sparse datasets MovieLens1M and Netflix, the noise-free subset from \name can improve more in performance. 
    
    \item In Table~\ref{tab:overall}, all denoising methods fail to enhance the performance of DeepFM, which may be attributed to two reasons.
    (i) For instance selection methods (Drop3~\cite{wilson2000reduction}, BDIS~\cite{Chen_Cao_Xing_Liang_2022}, LSBo~\cite{leyva2015three}, LSSm~\cite{leyva2015three}), they tend to compress the dataset, resulting in discarding excessive samples and missing critical information.
    (ii) For denoising models like T-CE and R-CE~\cite{wang2021denoising}, they improve the AUC score on a highly sparse dataset (Netflix, 0.02\%), yet are unable to provide benefit on a dense dataset. The reason is that T-CE and R-CE highly depend on the negative sampling strategy, which samples unobserved user-item interactions. However, for the highly-dense dataset as KuaiRec-Small (99.62\%), it would be hard to sample effective unobserved interactions.
    On the contrary, with the excellent exploring ability of the DRL structure, \name is compatible with all datasets of diverse densities.
\end{itemize}

\subsection{Transferability Test}
In practice, recommender system models are updated frequently, causing it time-consuming and costly to train a denoising framework for each model.
Therefore, we examine the transferability of \name in this section.

Specifically, based on DeepFM, after obtaining the noise-free dataset selected by \name, we directly use it to train other backbone recommendation models (e.g., FM~\cite{rendle2010factorization}, Wide \& Deep~\cite{cheng2016wide}, and DCN~\cite{wang2017deep}) and investigate their performances.
Results are presented in Table~\ref{tab:tran}, where `Normal' means training on the original data with noise, i.e., the `\textbf{w/o}' in Table \ref{tab:over1}.
We can observe that all backbone models' performances are enhanced by training on the noise-free dataset selected by \name. These experiments validate 
that \name's output has a strong transferability and the potential to be utilized in commercial recommender systems.

\subsection{Ablation Study}
We investigate the core components of \name through a range of experiments in this section. 
As we discussed in Section~\ref{subsec:retrain}, a vital part of \name is how to generate denoising results in the validation stage, i.e., adopting individual selection or top-$k$ selection.
Therefore, we design a variant of \name, named \textbf{\name-v}, to figure out the impact of different selection strategies.
The only difference compared to origin \name is that \name-v applies individual selection in policy validation while \name adopts top-k selection.

\begin{figure}[t]
    \includegraphics[width=0.77\linewidth]{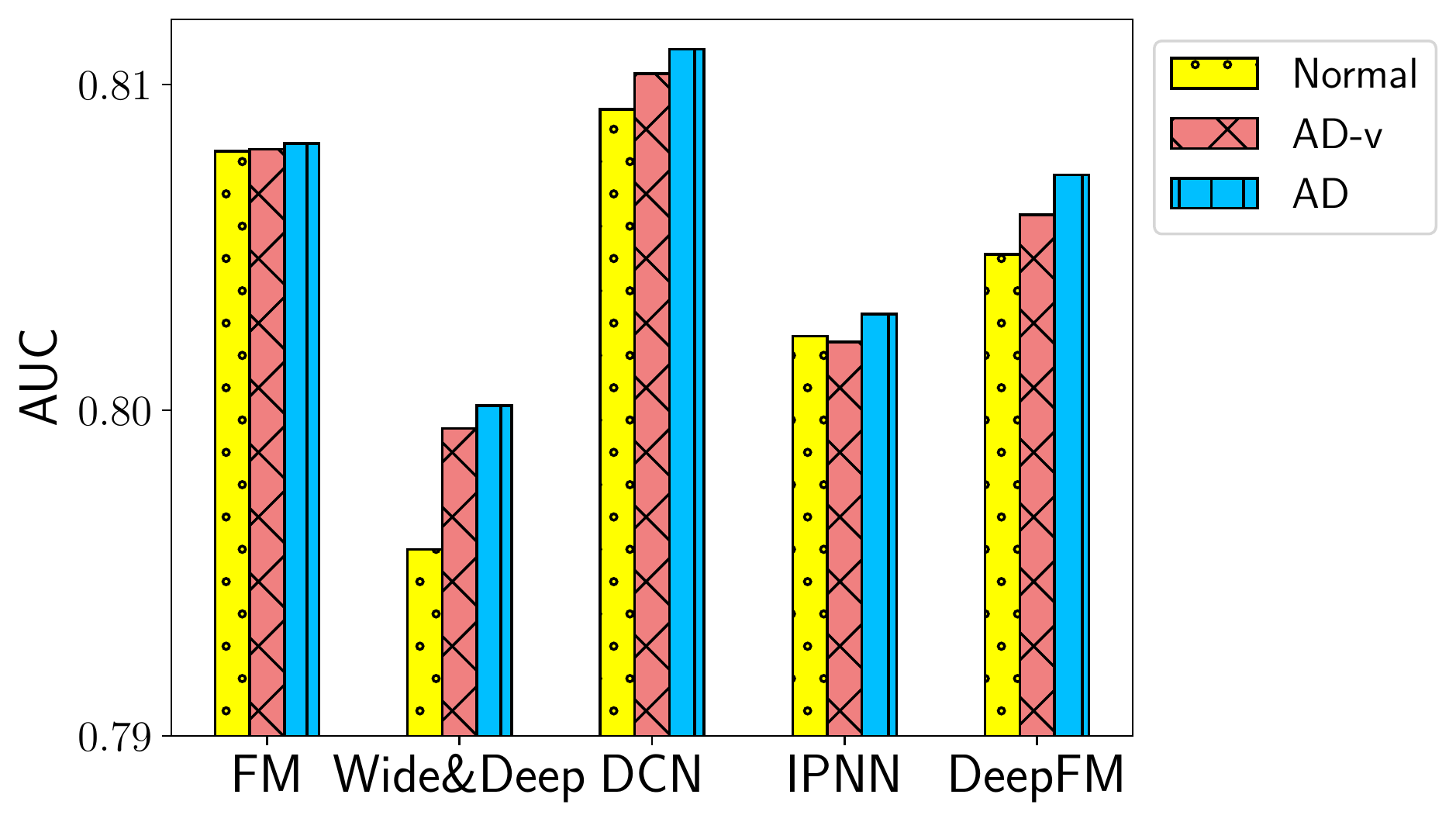}
    \Description{Compared the performance of Normal, AD-v and AD.}
    \vspace{-3mm}
    \caption{Ablation study on MovieLens1M. }
    \label{fig:ab}
    \vspace{-6mm}
\end{figure}

We test their AUC scores on MovieLens1M, and the results are illustrated in Figure~\ref{fig:ab}. 
It can be observed that \name (AD) outperforms \name-v (AD-v) on all models, and \name-v cannot generate noise-free datasets to improve IPNN performance.  
The reason is that \name-v focuses on individual scores while neglecting the global situation. Besides, \name could capture batch-wise global information to generate a robust policy by adopting a top-k selection strategy, which compares scores between the same batch of data. This experiment validates the rationality of our model design.

\subsection{Parameter Analysis}
\begin{figure}[t]
 \centering
 \Description{Show the performance changes on Warm-up epoch(left) and Selection ratio(right).}
 {\subfigure{\includegraphics[width=0.495\linewidth]{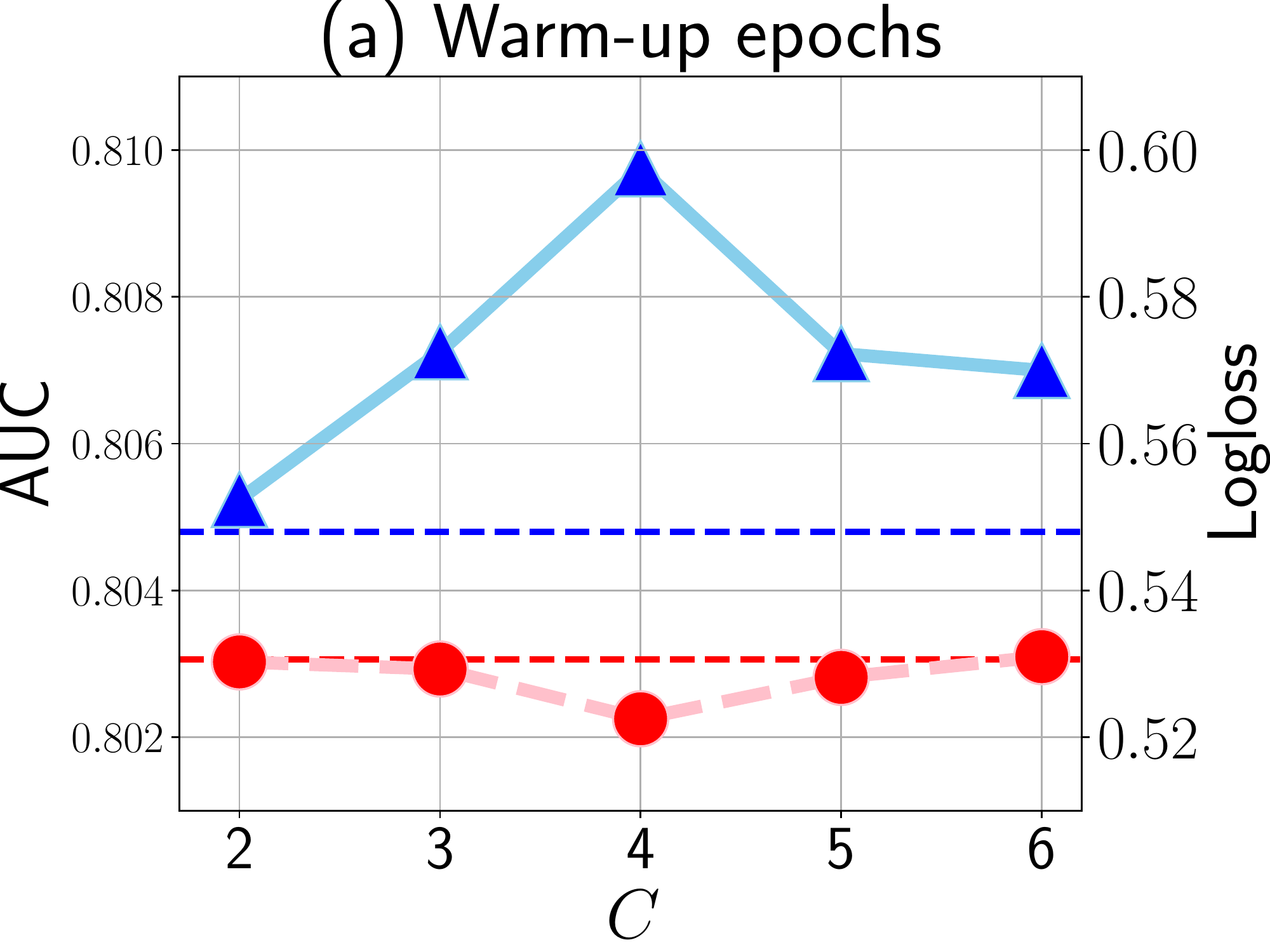}}}
 {\subfigure{\includegraphics[width=0.495\linewidth]{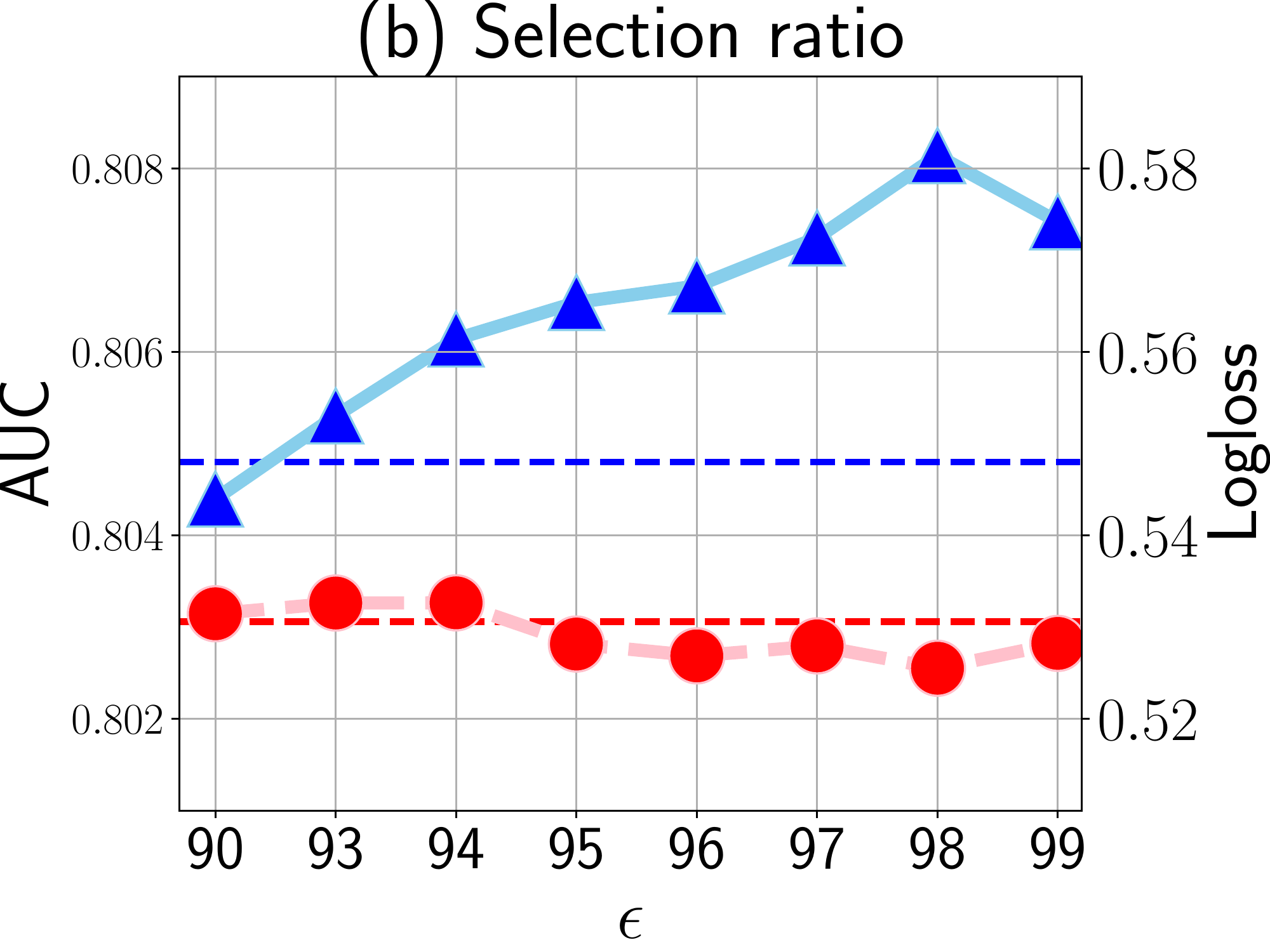}}}
 \vspace{-7mm}
 \caption{Parameter analysis on MovieLens1M.}
 \label{fig:param}
\vspace{-3mm}
\end{figure}

In this section, we conduct experiments on MovieLens1M with DeepFM to investigate the sensitivity of two critical hyperparameters: 
(i) the number of warm-up training epochs $C$; 
(ii) the number of selected samples $K$. 
The results are visualized in Figure~\ref{fig:param}, where two Y-axes are used to scale the AUC scores and logloss values, respectively.
The X-axis in Figure~\ref{fig:param} (a) represents the number of warm-up epochs $C$, and X-axis in Figure~\ref{fig:param} (b) is the selected ratio  $\epsilon = {K}/{M}$.
For both figures, blue lines illustrate the AUC score, and red lines are the Logloss value.
In addition, we also mark the `Normal' DeepFM performance as a dashed line for visual comparison (AUC $0.8048$, Logloss $0.5306$).
We can find that:
\begin{itemize}[leftmargin=*]
    \item With the changing hyperparameters, \name is superior to the Normal in most cases ($2\leq C\leq6$, $95\% \leq \epsilon \leq 99\%$), which can fully demonstrate its robustness.
    We achieve the best performance with the default setting (i.e., $C^*=4,\epsilon^*=98\%$).

    \item From Figure~\ref{fig:param} (a), the parameter $C$ directly influence the reward computation in Equation (\ref{equ:reward}). 
    For $C<C^*$, \name suffers from the randomness of too few warm-up epochs, i.e., the first term in Equation (\ref{equ:reward}), resulting in performance degradation. 
    However, a large warm-up epoch leads to a smooth average value, and the loss value of every batch, i.e., the last term in Equation (\ref{equ:reward}), have a more significant impact on the training process, leading to an unstable reward and impairing the denoising accuracy.
    
    \item As illustrated in Figure~\ref{fig:param} (b), it is crucial to make appropriate discards of instances.
    On one hand, dropping too many instances results in inferior performance with missing valuable information. 
    On the other hand, if we keep excessive instances, the model inevitably introduces noises and fails to get the best performance.
\end{itemize}

\subsection{Case Study}
We illustrate the effectiveness of \name with examples in this section. Specifically, we select a user (ID: 5051) from MovieLens1M and visualize part of the training set in Table~\ref{tab:casetr}, all test data instances in Table~\ref{tab:casets}. 
We present titles and genres of movies for the training set, together with the ground-truth click signals. 
For test data, we additionally exhibit the prediction result of DeepFM~\cite{guo2017deepfm} trained on original noisy data (Normal) and noise-free data (\name).

In Table~\ref{tab:casetr}, \name detects the noise \textbf{The Wizard of Oz} with a positive click signal. 
This movie is a children's and musical movie, while all the other movies the users prefer are mainly action, crime, and war movies.
By dropping this noise, \name contributes to correcting the prediction of DeepFM on \textbf{Wonderland} (from 0.46 to 0.57, in Table~\ref{tab:casets}), which is a documentary movie about war that the user usually enjoy. This case study also demonstrates the potential interpretability of \name.

\section{Related Work}
\label{sec:related_work}

In this section, we will briefly introduce the related works to our framework, i.e., instance selection and denoising.

\noindent\textbf{Instance Selection.}
Instance selection plays an important role in selecting the most predictive data instances to scale down the training set without performance degradation of predictive model~\cite{olvera2010review}. It can be divided into two groups, i.e., wrapper methods~\cite{hart1968condensed, chou2006generalized, wilson2000reduction, brighton2002advances}, and filter methods~\cite{riquelme2003finding, olvera2008prototype, raicharoen2005divide}, whose selection criterion is based on the model performance and a selection function, respectively. 
CNN~\cite{hart1968condensed} is one of the earliest proposed methods in the field of instance selection. It keeps the instances near the class border and drops the internal ones to generate an effective data subset.
DROP~\cite{wilson2000reduction}, a series of representative wrapper methods, relies on the instance associates, i.e., the k nearest neighbors on k-NN, to conduct the selection manner. 
LSSm and LSBo~\cite{leyva2015three} are methods based on local set~\cite{brighton2002advances}, i.e., the largest hypersphere that contains cases in the same class. LSSm defines and considers usefulness and harmfulness as measures to decide whether to remove an instance, while LSBo relies on the class borders to make the decision. In contrast to previous instance selection methods, \name uses a policy network to automatically distinguish noisy instances rather than a fixed selection rule, which can better discover the inner relationship in a highly dynamic data distribution.

\noindent\textbf{Denoising.}
The selection manner also works well in the fields of denoising, which can be categorized into selection-based method~\cite{ding2019sampler,gantner2012personalized,wang2021implicit,yu2020sampler}, along with the reweighting-based methods~\cite{wang2021denoising,wang2022learning}. The WBPR~\cite{gantner2012personalized} assigns higher selecting probabilities to the missing interactions of the popular items since it considers them to be the real negative instances. IR ~\cite{wang2021implicit} finds out the negative interactions and changes their label to revise the unreliable behaviors from users. R-CE and T-CE~\cite{wang2021denoising} consider that noisy instances would have high loss value, thus they dynamically assign lower weight to the high-loss instances as well as truncate the ones with weight lower than the threshold in two strategies, respectively.
However, the above models either suffer from selection bias or insufficient transferability, which are solved by \name with its DRL framework.

\begin{table}[t]
\small
\caption{Training dataset for user 5051.}
 \vspace{-3.3mm}
\label{tab:casetr}
\begin{tabular}{@{}ccc@{}}
\toprule
Title                             & Genres                      & Click      \\ \midrule
The Patriot           & Action|Drama|War            & 1          \\
Raiders of the Lost Ark   & Action|Adventure            & 1          \\
\textbf{The Wizard of Oz} & \textbf{Children's|Musical} & \textbf{1} \\
...                               & ...                         & ...        \\ \bottomrule
\end{tabular}
\vspace{-3mm}
\end{table}
\begin{table}[t]
\small
\caption{Test dataset for user 5051.}
\label{tab:casets}
 \vspace{-3.3mm}
\begin{tabular}{@{}ccccc@{}}
\toprule
Title                             & Genres               & Click & Normal/\name    \\ \midrule
The Extra-Terrestrial & Drama|Fantacy & 1     & 0.77/0.82 \\
Lawrence of Arabia         & Adventure|War        & 1     & 0.94/0.93 \\
\textbf{Wonderland}                 & \textbf{Documentary|War}      & \textbf{1 }    & \underline{\textbf{0.46/0.57}} \\ \bottomrule
\end{tabular}
\vspace{-3mm}
\end{table}
\section{Conclusion}
\label{sec:conclusion}
We propose a DRL-based instance denoising method, \name, to improve the recommendation performance of RS models by adaptively selecting the noise-free instances in every mini-batch of data and training the RS model with the selected noise-free subset. We design a two-phase optimization strategy to properly train and evaluate the proposed framework. The extensive experiments validate the effectiveness and compatibility of \name and its selected noise-free data subset. Further experiments prove the transferability of the noise-free subset, i.e., the noise-free subset selected with one RS model can transfer well to other state-of-the-art backbone RS models with significant performance improvement.

\subsection*{ACKNOWLEDGEMENTS}
This research was partially supported by APRC - CityU New Research Initiatives (No.9610565, Start-up Grant for New Faculty of City University of Hong Kong), SIRG - CityU Strategic Interdisciplinary Research Grant (No.7020046, No.7020074), HKIDS Early Career Research Grant (No.9360163), Huawei Innovation Research Program and Ant Group (CCF-Ant Research Fund).

\bibliographystyle{ACM-Reference-Format}
\bibliography{999Reference}

\end{document}